# Digital Democracy in the Age of Artificial Intelligence

Claudio Novelli[1], Giulia Sandri[2]

[1]Department of Legal Studies, University of Bologna, Via Zamboni, 27/29, 40126, Bologna, IT
[2]Centre d'Etude de la Vie Politique (CEVIPOL), Université libre de Bruxelles, Avenue Jeanne 44, 1050, Bruxelles, BE

**Abstract.** This chapter explores the influence of Artificial Intelligence (AI) on digital democracy, focusing on four main areas: citizenship, participation, representation, and the public sphere. It traces the evolution from electronic to virtual and network democracy, underscoring how each stage has broadened democratic engagement through technology. Focusing on digital citizenship, the chapter examines how AI can improve online engagement and promote ethical behaviour while posing privacy risks and fostering identity stereotyping. Regarding political participation, it highlights AI's dual role in mobilising civic actions and spreading misinformation. Regarding representation, AI's involvement in electoral processes can enhance voter registration, e-voting, and the efficiency of result tabulation but raises concerns regarding privacy and public trust. Also, AI's predictive capabilities shift the dynamics of political competition, posing ethical questions about manipulation and the legitimacy of democracy. Finally, the chapter examines how integrating AI and digital technologies can facilitate democratic political advocacy and personalised communication. However, this also comes with higher risks of misinformation and targeted propaganda.

## 1. Introduction

Digital democracy emerges from blending democratic governance with computer-mediated communications. Such integration often implies that it fundamentally transforms the political model, either enhancing or modifying its function. This integration leads to a significant qualitative shift in political engagement and governance.

Digital democracy represents just the latest in a series of paradigm shifts involving technology and politics. Berg and Hofmann have chronicle the evolution of digital democracy, illustrating how each phase has expanded the scope and depth of democratic engagement through technology (Berg and Hofmann 2021). They delineate three key stages: electronic democracy (e-democracy), virtual democracy, and network democracy. The inception of e-democracy in the 1980s used emerging technologies such as cable TV and bulletin board systems to create platforms for direct participation, like electronic town halls. This phase was followed by the 1990s era of virtual democracy, which leveraged the Internet to build global communities that supported techno-libertarian ideals and advocated for a form of democracy that transcended national borders. Entering the 2000s, the advent of Web 2.0 ushered in the network democracy era, characterised by massive participatory opportunities and the ambiguous role of digital platforms that now had to balance democratic ideals with commercial interests. This era also marked the rise of social media, highlighting a significant shift in where users consume and produce content, significantly enhancing their influence in the democratic public sphere (Berg and Hofmann 2021).

Ideally, social media platforms could enhance political deliberation and participation. By lowering barriers to accessing information, expressing opinions, and mobilising around essential issues, social media can empower citizens – as some empirical research suggests (Dzisah 2018; Jennings, Suzuki, and Hubbard 2021) – particularly those with lower incomes who may benefit from these accessible and inexpensive platforms (Congge et al. 2023). However, a significant concern lies in the potential for an unhealthy democratic debate, which could be further exacerbated by integrating (generative) Artificial Intelligence and its impact on communication and the overall social media environment.

In this chapter, our objectives are twofold. Firstly, we aim to outline the foundational components of digital democracy and their evolution over time. To do this, we will focus on four aspects: citizenship, participation, representation, and the public sphere.[1] Concurrently, we shall investigate how these elements might be transformed by the advent of AI technologies, evaluating whether such transformations indicate the onset of a new technological paradigm within digital democracy. It is important to note that many of the AI impacts we will discuss are speculative, as the application of AI in democratic processes is still in its early stages.

## 2. Digital Citizenship: from Individualised to Stereotyped Identities

Citizenship is a concept grounded in both political and legal domains. From a legal perspective, it denotes a status conferred upon an individual who both enjoys the rights and fulfils the duties associated with membership in a political community (Leydet 2023). Politically, citizenship goes beyond mere status, encompassing active involvement in that community's public life and institutions. This participation fosters a unique sense of identity for the citizen (Kymlicka and Norman 2000).

---

[1] The four dimensions of analysis are identified based on previous research on the topic, and particularly Jungherr's (2023) and Gilardi's (2022) pioneering studies.

Historically, citizenship was a stark legal and political discrimination within communities, evident in Roman and Medieval laws (Magnette 2005); during the Enlightenment, the notion of citizenship evolved to include a greater egalitarian dimension.

While citizenship itself is not a democratic invention, its evolution is deeply intertwined with the development of democracy. For example, the extension of citizenship rights — from property-owning men to all adults irrespective of property, race, or gender — has mirrored the progression of democratic practices (Saward 2008). The distinction between citizens and non-citizens has traditionally created significant differences in legal standing. However, this division has been considerably diminished by the advent of universal human rights and the constitutional innovations of the 20th century.

Providing a comprehensive history of citizenship is not feasible for this section. Instead, we aim to offer a basic overview underpinning the discourse on Digital Citizenship (DC), which evolves from broadening societal participation and responsibilities into the digital sphere. The concept itself is neither fixed nor clear-cut. Definitions often swing between overly broad and overly prescriptive, suggesting a blend with the ideal of 'good' DC and emphasising responsible technology use (Ribble, Bailey, and Ross 2004). For the same reason, DC frequently intersects with education and media literacy in scholarly discussions (von Gillern, Gleason, and Hutchison 2022). Sometimes, it is argued that the digital realm has fundamentally altered the nature of citizenship, making it more individualised — particularly in electoral contexts — where individual identity is prioritised over collective identity (Ceccarini 2021).

However, for a more abstract and descriptive approach, DC can be defined as the capacity to participate in a political community via information technologies (Mossberger, Tolbert, and Mcneal 2007, 1).

Moonsun Choi identifies the following four integral dimensions of DC (Choi 2016):

(1) DC as Ethics emphasises the importance of engaging appropriately, safely, ethically, and responsibly online, recognising virtual communities as platforms where individuals frequently interact and communicate.

(2) DC as Media and Information Literacy extends beyond the ethical use of digital technologies to include the ability to access, use, create, and critically evaluate information, underscoring the necessity to bridge the digital divide and ensure universal internet access.

(3) DC as Participation/Engagement explores the role of the Internet in facilitating both broad and personalised forms of political, socio-economic, and cultural participation, highlighting activities ranging from e-voting to more personal cultural interactions online.

(4) DC as Critical Resistance, while overlapping with participation, specifically calls for transformative actions that challenge existing power structures and promote social justice, often through innovative, decentralised methods.

Together, these categories construct a layered understanding of digital citizenship, reflecting its normative and active engagement aspects in the digital age.

## 2.1. The role of AI technologies

In this context, AI technologies are significantly transforming DC, influencing it in both beneficial and problematic ways. When viewed through its normative dimension, AI can enhance DC by fostering safer online environments, e.g., monitoring and moderation systems that identify and address harmful content, including misinformation, cyberbullying, and illegal activities. In this sense, AI algorithms are employed by social media platforms to detect and remove content that breaches terms of service or raises ethical concerns (Zarnoufi, Boutbi, and Abik 2020; Nirban et al. 2023). However, there are increasing worries about AI's potential to undermine aspects of DC, such as privacy, surveillance, and the possibility of censorship (Monteith et al. 2024). These concerns are particularly pronounced with Generative AI, such as large language models (ChatGPT, Gemini, or Claude) that are widely accessible to the public and possess significant capabilities for creating damaging content and spreading misinformation (Xu, Fan, and Kankanhalli 2023; Gabriel et al. 2024; Novelli, Casolari, et al. 2024).

Moreover, AI is changing digital awareness by delivering personalised educational content that adapts to an individual's learning pace and style. High-quality, personalised AI-driven education can enhance digital literacy, equipping users with the skills to evaluate online information critically and potentially bridging educational gaps. Yet, personalisation comes with risks of manipulating perceptions or reinforcing biases, which clashes with a liberal reading of awareness.

Against this background, AI technologies significantly impact the individualisation of citizenship, an evolution shaped by the digital era, albeit in a distinct manner. Through data analysis and algorithmic processing, AI emphasises the individual dimension of citizenship within broader patterns of statistical correspondence. By analysing vast amounts of data, AI can identify patterns and predict behaviours at an individual level. However, these technologies often depend on generalised data models that can result in stereotyping (Kotek, Dockum, and Sun 2023; Garcí-a-Ull and Melero-Lázaro 2023). For example, in political advertising and content recommendation systems, algorithmic decision-making categorises individuals into specific demographic or psychographic groups (Papakyriakopoulos et al. 2022). This practice tends to simplify the complex spectrum of individual identities into broader, sometimes inaccurate and biased, stereotypes. Stereotypes may be based on demographic information such as age, gender, ethnicity, or location; socioeconomic biases that pigeonhole individuals by economic status; behavioural categorisations from online activity; cultural stereotypes linked to specific ethnic groups; and engagement levels inferred from digital interactions.

Stereotypes can distort the experience of digital citizenship and create feedback loops where they are continually reinforced (Martínez et al. 2024). As individuals interact with content that aligns with AI's categorisation, their responses further validate and strengthen the algorithm's predictions, potentially entrenching these patterns further. These loops can hinder personal growth and the evolution of individual identities within the digital realm, compromising the exercise of aware citizenship and the potential for transformative citizenship.

We have several strategies to mitigate these risks, even though AI system providers and platforms may not always find them profitable. For instance, algorithmic fairness models can prevent or compensate for such biases. They analyse and measure unfairness in model predictions by comparing privileged and unprivileged demographic groups using different fairness definitions, such as equalised odds and disparate impact. Fairness assessment begins with exploratory data analysis to identify potential sources of bias. To correct and prevent unfairness, various quantitative approaches can be applied at different stages of model development, including pre-processing, in-processing, and post-processing methods (Pessach and Shmueli 2023).

## 3. Participation: Civic Engagement and Digital Platforms

One feature of digital citizenship that deserves specific attention is political participation. This refers to how citizens actively engage in the social and political process to influence decisions affecting their community.

Traditionally, political science focused on electoral participation (Verba and Nie 1987) as the primary way for citizens to be heard. Voting turnout was the most common measure of citizen engagement in the US (Ekman and Amnå 2012). However, the definition has broadened over time to include non-voting actions that influence community life. Demonstrations, strikes, boycotts, and other forms of protest are now considered political participation (J. W. van Deth 2014). This acknowledges that citizen actions can target various actors, including political figures, social institutions, media outlets, and economic forces (Norris 2002).

Digital technologies, especially those that enable social interaction, can create new avenues for political engagement. Online interactions can influence voting behaviour, participation in political movements, and even the motivations behind different forms of participation in many ways (Koc-Michalska and Lilleker 2017; Theocharis et al. 2023). Take social media, for example; it has empowered citizens to launch and organise boycott campaigns against political figures or policies with unprecedented ease. However, this double-edged sword can also have negative consequences. The tools that connect us can spread propaganda, exacerbate polarisation, and promote a superficial kind of engagement, namely "clicktivism" (Borbáth, Hutter, and Leininger 2023).

While mapping the impact of digital technologies on political participation is a complex endeavour, partly due to the technological heterogeneity of social platforms (Theocharis et al. 2023), valuable frameworks can aid our understanding. Teorell et al. (2007) proposed a five-dimensional typology of political participation that remains relevant even if initially developed for a broader understanding of the concept (Teorell, Torcal, and Montero 2007).[2] We are analysing their five dimensions through the lens of technological affordances: "the actions and uses that a technology makes qualitatively easier or possible when compared to prior like technologies" (Earl and Kimport 2011, 33).

(a) *Voting.* Voting exemplifies a representational mode of participation, where citizens exercise their influence by casting votes for candidates or parties in elections, thereby shaping government policies. It operates as an exit-based mechanism, akin to how consumers choose to purchase or abstain from buying a product based on its quality; if the quality of governance deteriorates, some voters may opt not to vote for the incumbent party or candidate (Teorell, Torcal, and Montero 2007).

Although e-voting technology has been under development for over three decades, its widespread implementation remains elusive in most systems. The primary challenges lie in providing a secure solution for each electoral stage and gaining voters' trust in using it (Wang et al. 2017). In countries where e-voting is practised, such as Bulgaria and Estonia (Tsahkna 2013; Tsareva 2020), one of the most significant potential impacts of digital technologies on voting is the increased accessibility and convenience they offer (see *infra* section 4). E-voting systems and online voter registration platforms enable citizens to participate in elections without a physical location or time constraints (Smith and Clark 2005). Furthermore, online resources can enhance voter literacy by providing detailed information about candidates, their platforms, and their positions on various issues, informing voters' choices. However, this also raises concerns about the spread of disinformation, particularly among older generations (Goldstein 2020).

[2] There are some similarities with Verba and Nie's four modes classification: voting, campaign activity, citizen-initiated contacts, and cooperative participation (Verba, Schlozman, and Brady 1995).

AI technologies are increasingly being explored for various applications within the voting process. For instance, research is ongoing into AI algorithms designed to monitor voting patterns in real-time to detect irregularities and unauthorized access, verify voter identity and eligibility, and cross-check voter rolls, thereby adding a layer of security (Tamilselvi, Manimaran, and Inunganbi 2023). AI is also being integrated into Voting Advice Applications (VAAs), which provide personalised voting recommendations based on users' policy preferences. AI can automate the selection of salient issues for formulating questions in VAAs, enhancing their relevance and accuracy (Buryakov et al. 2022; Gemenis 2024). Additionally, AI-based image processing systems are being considered for automatic ballot recognition, capable of efficiently counting and summarising votes and displaying results (Zhao, Yan, and Wang 2021).

(b) *Consumer Participation*. Consumer participation is a form of political participation that operates outside of formal political structures like voting. Instead, it leverages economic actions to influence political outcomes. Akin to market behaviour, individuals use their economic power to support or oppose specific policies or practices. For example, donating to political causes or candidates can amplify particular voices, while boycotting products or services can signal dissatisfaction and prompt change. Like voting, consumer participation is an exit-based mechanism where citizens can withdraw their support or patronage as a form of protest (Boulianne 2022).

Digital technologies have facilitated consumer participation in several ways. Online platforms and social media enable individuals to raise awareness, mobilise support, and solicit donations for political causes or candidates they endorse. Crowdfunding campaigns and online petitions, significantly when enhanced with features like identity disclosure and project tracking (Kim, Por, and Yang 2017) [3], have given more prominence to voices that might have been marginalised through traditional channels. Websites like

---

[3] This is especially has been documented for non-political crowdfunding initiatives, but we may expect similar tendencies for political ones.

Change.org allow users to start and sign petitions that can pressure companies or governments to revise their policies.

With the opposite aim, the internet and social media have made it easier to organise and coordinate boycotts against companies or products whose practices or policies are deemed unacceptable (Kelm and Dohle 2018). Platforms and apps like Buycott and DoneGood provide information on corporate practices, empowering consumers to align their purchases with their political and ethical values, thus exerting economic pressure for change.

In this context, AI technologies can further assist consumers in making politically motivated purchasing decisions. For instance, AI algorithms, such as recommender systems, can analyse an individual's purchase history, values, and preferences to suggest products, services, or brands that align with their ethical or political beliefs. Additionally, AI can optimise the targeting, messaging, and delivery of consumer activism campaigns, like online petitions or boycott initiatives. By analysing user engagement, demographics, and behaviour data, AI can tailor campaigns for maximum impact and effectively reach the right audiences. On the other hand, AI algorithms also pose risks to this dimension of political participation, as they can manipulate consumer choices through tactics such as dark patterns (de Marcellis-Warin et al. 2022).

(c) *Party Activity.* Party activity is a representational form of political participation, consisting of direct engagement with political parties. Party activity relies on a voice-based mechanism characterised by efforts to influence political outcomes from within. Party members and activists seek to shape party platforms, policies, and candidate selections through internal processes rather than leaving the party. This includes various activities, such as campaigning for party candidates, becoming a party member, volunteering, and donating to the party. These activities help organise and mobilise support and develop and communicate political agendas.

Political parties use online platforms, websites, and social media channels to engage with members, volunteers, and supporters. These digital tools facilitate communication, dissemination of information, mobilisation efforts, and coordination of activities. Increasingly, party members participate in online discussions, provide feedback, and stay updated on party initiatives. Additionally, some political parties have adopted digital platforms for internal decision-making processes, such as voting on party policies, platforms, or candidate selections (Barberà et al.

2021; Sandri et al. 2024). Examples include the Italian Five Star Movement (Movimento 5 Stelle), which uses the Rousseau platform; the Pirate Party in Germany, which uses the LiquidFeedback platform; and Podemos in Spain (now called Sumar), which employs the Participa platform. These digital systems allow members to propose laws, participate in discussions, and vote on various issues, including candidate selections and party policies.[4]

In this context, AI technologies can streamline the analysis of large datasets, including social media, voter information, and demographic data. This enables political parties to identify potential supporters, understand their preferences, and tailor their campaigning and messaging strategies accordingly. Additionally, these data sources, which differ from self-reported information, can improve internal party democracy by providing more transparent and accurate evaluations of the inclusiveness of decision-making processes (Novelli, Formisano, et al. 2024).

Furthermore, AI can be integrated into digital platforms to enhance various processes, such as moderating discussions, analysing and summarising proposals through Natural Language Processing (NLP) (Hadfi et al. 2021), and creating visual representations of arguments and counterarguments made during online deliberations (Zhang et al. 2023). NLP can also detect logical fallacies (Sourati et al. 2023), which is particularly beneficial for political debates and deliberative practices by enabling participants to focus on more robust arguments and identify the weaknesses in others' reasoning.

Techniques like sentiment analysis can also be applied to online discussions and feedback from party members to gauge their reactions, opinions, and levels of support or opposition regarding specific proposals or decisions (Novelli, Formisano, et al. 2024).

(d) *Protest Activity*. Protest activity is an extra-representational form of participation outside formal political structures and often directly challenges existing policies or power structures. Unlike exit-based mechanisms, protests represent a collective vocalisation of grievances and demands for change. Demonstrations and strikes are organised expressions of dissent and are intended to draw public attention and force political concessions. Protests can take many forms, including marches, rallies, sit-ins, and strikes. They often involve mobilising large numbers of people to physically occupy public spaces or disrupt normal activities to draw attention to specific issues.

---

[4] Literature shows a correspondence between the agendas of the Five Star Movement's elected representatives and the priorities of their platform's members (Mosca and Vittori 2023).

Digital technologies have increasingly facilitated the dissent aspect of political participation, similar to how they enhanced consumer participation. Social media platforms, messaging apps, and dedicated online tools are used to coordinate and mobilise protestors. These technologies allow organisers to disseminate information about planned demonstrations, share logistics, and rally support (Lee and Chan 2016). For example, digital technologies have played a crucial role in organising protests against traditional media narratives, as seen with the Gilet Jaunes movement (Baisnée et al. 2022). In countries with strict media censorship or internet restrictions, protesters often use virtual private networks (VPNs), encrypted messaging apps, and other digital tools to bypass censorship and maintain communication channels. Moreover, smartphones and other digital devices capture photos, videos, and live streams of protest events, providing documentation and evidence of incidents or confrontations with authorities. Despite these advancements, authoritarian regimes have developed sophisticated censorship strategies to control the digital environment (Feldstein 2021; Stoycheff, Burgess, and Martucci 2020; Kawerau, Weidmann, and Dainotti 2023).

On the one hand, AI technologies can significantly enhance the organisation and mobilisation of protests, such as by improving communication among protesters, particularly in environments with heavy censorship or surveillance, through AI-driven encryption technologies (Curzon et al. 2021). On the other hand, AI provides tools for authorities to monitor, control, and suppress dissent. For example, AI-powered facial recognition technology used by governments can analyse video footage from public spaces to identify individuals participating in protests, leading to arrests and other punitive actions. China's extensive use of AI-driven surveillance has been documented in monitoring and suppressing Uighur Muslims and political dissidents (Feldstein 2019, 45; Leibold 2020). Additionally, authorities may use AI to monitor social media platforms for signs of dissent, identify protest organisers, and automatically censor content. It is also important to note that censorship may originate from social platforms themselves rather than political authorities (Cobbe 2021; Bradford 2023).

(e) *Contact Activity*. Contact activity spans both representational and extra-representational channels of political communication.

This dimension involves citizens directly communicating their concerns, opinions and demands to political actors and institutions. It includes writing letters, sending emails, making phone calls, and meeting with politicians or civil servants. Contacting officials and organisations allows citizens to voice their concerns and influence decision-making processes. However, unlike protests, it often involves more formal and institutionalised engagement methods.

Digital technologies have lowered barriers to political participation, enabling more people to engage in contact activities. E-government services, such as websites for public service announcements, have streamlined communication between citizens and government officials, resulting in quicker responses and more efficient inquiry handling (Manoharan and Ingrams 2018). While digital communication's immediacy can lead to more spontaneous interactions with political figures, this shift brings challenges, e.g., the spread of misinformation and more reactive, less thoughtful communication.

This also relates to civic tech, defined as technology used to enhance democratic participation (Gilman, 2017: 745). Civic tech fosters a bottom-up approach to collaborative governance that transforms the citizen-state relationship through digital tools, with participatory budgeting being a prime example of this engagement (Barros and Sampaio, 2016: 296).

AI can significantly impact this dimension of political participation by facilitating more efficient and seamless communication between citizens and government officials or political organisations. For example, chatbots and automated response systems can handle routine inquiries from citizens, identify common concerns, and prioritise responses based on urgency and relevance. Research on this kind of application has been conducted on using NLP to assist Madrid citizens through the "Decide Madrid" Consultative platform, which enables citizens to post proposals for policies they would like the city council to adopt (Arana-Catania et al. 2021).

Moreover, by automating the processing of large volumes of communication, AI can help distinguish between meaningful interactions and less relevant noise (e.g., through spam filters or deduplication) (Mahesh et al. 2020). This ensures that political actors are not overwhelmed by the sheer volume of digital communications and can focus on substantive issues.

## 4. Representation: Digital and AI Technologies in Modern Electoral Processes

The integration of digital technologies has become pervasive throughout all phases of the electoral process, whether at the local, regional or national level, fundamentally transforming how elections are conducted and managed. This technological incorporation spans from geographical tools to the transmission of results, ensuring efficiency, accuracy, and transparency in the electoral process (Garnett and James 2020). Digital technologies, using geographical analysis software, begin to play a crucial role even before the actual voting process starts. Technologies such as Geographic Information Systems (GIS) are employed to delineate district borders and determine the optimal locations for polling stations. For example, GIS technology can analyse demographic data and geographic factors to ensure that polling stations are accessible to most of the electorate, thereby enhancing voter turnout and participation (Kim 2020).

The voter registration phase of the electoral process also benefits significantly from digital technologies. Digital systems are used to create and maintain accurate voter rolls, which are essential for the legitimacy of the electoral process. For instance, biometric registration systems, which use fingerprints or facial recognition, ensure that each voter is uniquely identified and registered, reducing cases of fraud and duplication. Studies have shown how biometric technology in voter registration has been successfully implemented in countries like Ghana, Chad and Kenya, leading to more reliable voter rolls (Jacobsen 2020; Debos 2021).

Moreover, in several countries, especially in large, populous democracies such as India, digital technologies facilitate the registration of parties and candidates by providing online platforms for submitting and verifying required documents. This digital approach streamlines the process, making it more efficient and less prone to human error.

During the actual voting phase, digital technologies are used through various means, such as ballot scanners, electronic voting (e-voting), and internet voting (i-voting). Ballot scanners are used to quickly and accurately count paper ballots, reducing the time and potential errors associated with manual counting. As used in India, electronic voting machines provide a secure and efficient way for voters to cast their ballots (Rajeshwari 2020). Internet voting, while still under scrutiny for security concerns, has been implemented in countries like Estonia, allowing citizens to vote remotely, increasing voter participation and convenience (Ehin et al. 2022). The tabulation of results is another critical phase where digital technologies ensure accuracy and transparency. Digital systems aggregate voting data from various polling stations, facilitating real-time updates and quick dissemination of provisional results. For example, using centralised databases and secure servers in the United States has allowed for faster and more reliable vote counting and result tabulation.

Finally, digital technologies are pivotal in the transmission of results. Secure electronic transmission systems ensure that results are conveyed quickly and without tampering from local polling stations to central counting centres. For instance, the Election Commission of India has employed the Electronic Voting Machine (EVM) and Voter Verifiable Paper Audit Trail (VVPAT) systems for more than a decade to ensure the secure and swift transmission of voting data, enhancing the integrity of the electoral process.

Digital technologies for electoral processes nowadays are primarily used in Internet voting systems. Internet voting allows people to cast their ballots remotely and without supervision over the Internet via a connected device (a computer or a smartphone), eliminating the need to visit a polling station. It is distinct from electronic voting, which often includes the use of direct-recording electronic voting machines (DRE) usually installed in polling stations and thus physically supervised by some electoral management authority. More generally, electronic voting systems rely on electronic technology for their functionality and use electronic means to either aid or manage the casting and counting of ballots (Krimmer and Barrat 2022). E-voting may use standalone electronic voting machines or computers connected to the Internet, depending on the implementation. Internet voting is thus a specific subtype of electronic voting systems.

Nowadays, electronic voting is used nationwide in 4 countries: Brazil, India, Venezuela (on-site DRE) and Estonia (optional internet voting). Electronic voting nationwide means it is offered to all voters in the country. Additionally, electronic voting is used by a majority of voters (but not all, so in different types of elections) in 4 countries: Belgium, Brazil, India, and Venezuela (on-site electronic voting). Estonia is one of the very few countries in the world using Internet voting. Overall, there are 13 main countries using internet voting, often with combined other methods of electronic voting, according to the type of election: Canada, US, Mexico, Pakistan, India (since 2021), Japan, Australia, New Zealand, Oman, United Arab Emirates, Armenia, France and Estonia[5].

Internet voting is often presented by its advocates as a potential remedy for many issues arising in electoral processes (for an overview, see Germann and Serdült, 2017). For instance, internet voting is considered an instrument for combating abstention by encouraging the participation of groups that traditionally are less likely to vote (such as young people) and by facilitating access to the ballot box for specific voter groups (expats, people with disabilities, etc.). It is also praised for reducing election management costs (eliminating polling stations, speeding up the counting process, etc.).

However, in a representative democracy, the electoral process must meet various conditions in terms of both democratic norms and technical standards. The security of the ballot and its verification process must be ensured (Fitzpatrick and Jost 2022), as well as the management, ownership, hosting and control of the data resulting from the internet voting process. It is because of these legal and technical vulnerabilities that several countries have abandoned this voting option (the Netherlands) or ruled it out (Great Britain) (von Nostitz et al. 2021). As part of the process of digital transformation of politics and society, internet voting raises more general questions about citizens' relationship with the electoral process. In addition to fears about the security of such online polls, political opposition to this instrument - particularly at the extremes of the political spectrum - is developing (Lust, 2015). As a result, the political class may be divided over the introduction of Internet voting.

AI can significantly impact this dimension of political representation by facilitating more efficient, safe and cost-cutting electoral processes. In the USA, the Electronic Registration Information Center (ERIC) uses machine learning to maintain voter rolls (Electronic Registration Information Center, 2024). Biometric verification systems employing deep learning models are used or tested in many countries (Wolf et al., 2017), and signature matching tools are standard in the USA. Advances in deep learning and transformer architecture have enabled the analysis of extensive unstructured data, including text, images, and video. These developments, along with large datasets and computational power, have led to generalised models, although they often lack interpretability (Juneja 2024).

LLMs help Election Management Bodies (EMBs) analyse and summarise complex texts. Other generative AI models handle diverse outputs like video and audio. AI techniques such as graph neural networks and boosting methods offer new analytical avenues. Global trials and evolving AI guidelines will influence EMB resources and capabilities, requiring careful consideration of AI's trade-offs and impacts on electoral processes. More specifically, AI can play a significant role in various stages of the electoral process, particularly in campaign and media monitoring, voting, and tabulating results. The application of AI in election management offers promising potential but also presents significant challenges and risks (Juneja 2024). One major use case for EMBs involves employing LLMs and graph neural networks on social media platforms to detect and summarise common misinformation about elections. This requires careful planning, contextual understanding, language training, and identifying key misinformation platforms (Dhiman et al., 2023). However, excessive reliance on AI could result in missing critical issues, especially on private messaging platforms[6].

---

[6] https://www.oas.org/en/iachr/expression/docs/publications/internet_2016_eng.pdf
, accessed on 4 July 2024.

EMBs are tasked with monitoring communications from campaigns, political groups, and media organisations, where AI can help detect violations of mandated campaign silence periods. LLMs can be fine-tuned to flag content violations by analysing social media inputs from political campaigns. Yet, overreliance on AI tools may result in missed violations and potentially discriminatory outcomes due to poorly designed training data (Juneja 2024).

AI can also enhance voting operations, such as voter identification verification through document and address verification models. However, traditional methods like barcode scanning might be more effective in some contexts. AI could inadvertently disenfranchise voters without improving security in areas with low voter fraud or identification issues (Hajnal et al. 2018). Additionally, biometric recognition for voter verification could improve security and risk discriminatory disenfranchisement and data privacy concerns (Padmanabhan et al. 2023).

AI's role extends to monitoring polling places by detecting incidents reported on social media, although overreliance might miss critical information. Governments' mass data collection for these purposes raises privacy and human rights concerns. For vote tabulation and analysis, AI can improve optical scanning systems for counting ballots. However, low failure rates of AI systems still pose significant risks to electoral integrity and public trust (Zhao et al., 2023). AI can also help EMBs by analysing real-time voter turnout and detecting anomalies, similar to post-electoral audits. AI models can predict and monitor turnout metrics, flagging potential issues for investigation. However, inaccuracies in pre-election simulations could lead to false positives and misallocated resources. For new voting methods like internet voting, AI can enhance security through facial recognition for identification, bot activity detection, and cybersecurity threat assessment. AI can also test online voting systems for vulnerabilities (Juneja 2024).

General-purpose AI tools like ChatGPT, Microsoft's Copilot, and Google's Gemini can improve EMB productivity by summarising documents, drafting content, and assisting with code. However, reliability and security concerns remain, as these tools might generate inaccurate information or cause data leaks. GenAI companies are developing government-ready versions of these tools to mitigate risks, and EMBs should emphasise human oversight and thorough fact-checking[7].

Finally, AI can predict electoral behaviour and reduce political uncertainty. Democracies rely on elections to channel and manage political conflict, allowing political parties to gain power within a given institutional framework. Thus, each party must believe in a real opportunity to win future elections, creating a system of organised uncertainty (Przeworski, 1991, p. 12–13). The application of AI promises to offset this organised uncertainty by predicting electoral outcomes. Barack Obama's campaigns were the first to use data-driven models to predict voter behaviour (Obama predicts, see: Hersh 2015). All ML/AI models allow using available information to infer unknown outcomes. This can occur individually, using survey attitudes, sentiment analysis, or documented actions to predict future behaviours like voting or donating. On a system level, aggregate data such as economic conditions or approval ratings can predict election outcomes without modelling individual behaviour. Thus, AI might reduce uncertainty about election results (Sumi 2021; Kefford et al. 2023).

[7] https://www.axios.com/2024/03/29/congress-house-strict-ban-microsoft-copilot-staffers accessed on 4 July 2024.

However, predicting individual voting behaviour remains limited. Committed partisans' behaviour can be predicted with some probability, especially in two-party systems, but predicting the actions of less politically involved individuals is much harder. Voter behaviour varies, and vote choices are often unavailable to modellers, making AI less suited for this task. As a result, some election outcome uncertainty persists, though data-driven models can still provide campaigners with competitive advantages by predicting probabilities of electoral participation or financial contributions at the individual level (Kreiss 2016).

Firms and governments might also use AI to predict election outcomes or the electorate's mood, potentially intervening in the process. Despite these efforts, the public's perception of AI's capabilities could undermine election legitimacy and provide a pretext for challenging results (Brauner et al. 2023; König 2023). The Cambridge Analytica scandal during Brexit and the 2016 U.S. presidential election exemplifies public concerns about AI's role in election manipulation, even if its actual impact was minimal. The direct influence of AI on elections seems minimal because of the scarcity of the observed activity, namely voting. Although AI can offer some advantages in campaign strategies, these are unlikely to result in a lasting power shift. A more significant issue is the public's perception that AI can eliminate electoral uncertainty, potentially eroding trust in the electoral process and its results (Jungherr 2023).

## 5. Public Sphere and Political Advocacy

The public sphere is a space in social life where public opinion can be formed and is accessible to all citizens. It allows private individuals to come together to discuss and address societal problems, influencing political action (Habermas, Lennox, and Lennox 1974; Habermas 1989). Digital technologies have significantly transformed the public sphere in advanced democracies. In today's rapidly evolving political landscape, digital technologies are an effective tool for structuring advocacy, democratising access to political platforms and fostering inclusivity. By enhancing outreach efforts and optimising campaign organisation, digital technologies empower diverse voices and level the playing field for political contenders of all sizes (Dennis and Hall 2020).

Moreover, digital technologies have the capacity to transform how political actors engage with voters, facilitating personalised communication strategies that resonate with individual concerns and priorities. This tailored approach fosters greater civic participation and cultivates a more informed electorate, thereby strengthening the foundations of democracy (Jungherr et al. 2020). However, legitimate concerns surround digital technologies' potential to amplify negative political campaigning and communication. From the proliferation of misinformation to deepfakes, tangible risks must be addressed to safeguard the integrity of elections and protect vulnerable communities.

Online advocacy campaigns use digital tools and social media platforms to mobilise support, raise awareness, and drive social and political change. These campaigns have become significant due to their ability to reach large, diverse audiences quickly and cost-effectively. Social media platforms such as Facebook, Twitter, TikTok and Instagram play crucial roles in these efforts, disseminating messages, engaging with supporters, and coordinating actions (Santini et al. 2022). Successful examples include the #MeToo and BlackLivesMatter movements, the Climate Strike movement and the Arab Spring, which used digital technologies to amplify voices and organise protests (Bennett and Segerberg 2023). The benefits of online advocacy include increased reach and engagement, allowing campaigns to mobilise supporters and influence public opinion effectively. However, these campaigns also face challenges, such as the risk of misinformation and spreading false information, which can undermine credibility and trust (Kalsnes 2023).

Moreover, the integration of digital technologies has profoundly impacted also political advertising, with online political marketing and e-campaigning becoming essential elements of election campaigns worldwide. Initially, online political tools catered to small audiences, but they have evolved into sophisticated components of hybrid media systems, enabling data-driven campaigns (Chadwick 2017). Data-driven campaigns and propaganda leverage advanced data analytics and targeted advertising to influence public opinion and political outcomes. These campaigns analyse vast amounts of data to tailor messages to specific audience segments, optimising engagement and effectiveness (Dommett et al. 2024).

Data analytics allows for precisely targeting ads based on demographic, psychographic, and behavioural data, enhancing the ability to sway opinions and mobilize supporters. This evolution began with campaign websites in the mid-1990s and progressed to individual-centred campaigns leveraging social media and smartphone apps. Notable milestones include using campaign websites in 1996, Howard Dean's fundraising weblog in 2004, and Barack Obama's sophisticated voter database in 2008 and 2012 (Kreiss, 2016). Donald Trump's 2016 and 2020 campaigns further advanced this trend by using social media data and heavily investing in targeted ads (Jungherr et al., 2020).

While expanding traditional campaign formats, online political advertising has been scrutinised for its democratic impact, competition dynamics, and communication changes. Despite initial radical expectations, online political advertising has augmented rather than replaced traditional campaigning methods. It has strengthened campaign formats through voter data usage but only moderately influences citizen participation (Boulianne, 2020). The anticipated consequences of rising online political advertising in hybrid media systems were encapsulated in three hypotheses: erosion, equalisation, and normalisation (Ward and Gibson, 2009). Empirical findings favour normalisation, with resource-rich political actors more likely to succeed. Campaign functions have not changed but are more interrelated due to online communication (Roemmele and Gibson 2020).

However, the lack of transparency and regulation of data-driven campaigns and online advertising has sparked concerns (Fowler et al. 2021). Traditional political ads have minimal effects, extending to online ads, which may influence media narratives rather than directly persuading voters (Kalla and Broockman 2018). The effectiveness of ad targeting varies, with studies indicating significant improvements in commercial contexts and highlighting biases due to social media's optimisation algorithms. The Cambridge Analytica scandal also exemplifies using personal data from social media to craft targeted political advertisements (Hinds et al. 2020). Ethical considerations include significant privacy concerns and data security issues, as personal data is often collected and used without explicit consent. The ethical implications of such targeted propaganda raise questions about manipulation and the potential erosion of democratic processes (Zuboff, 2019). Ensuring transparency and accountability in data usage is crucial to mitigate these risks.

One of the main risks and negative externalities of online political advertising and data-driven campaigns is spreading misinformation. Misinformation is false or inaccurate information spread unintentionally or without intent to deceive. Unlike disinformation, which is deliberate, misinformation can be unknowingly disseminated by those who believe it to be true. It can spread through various channels like social media, news outlets, and personal communication and can take forms such as rumours, conspiracy theories, and misleading information (Kalsnes 2023). Political disinformation aims specifically to shape public opinion, influence political outcomes, or gain an advantage over opponents. It spreads through channels, including social media, news outlets, and propaganda, often creating confusion, distrust, and division. Tactics include emotional or biased appeals, media manipulation, fake social media accounts, bots, and propaganda. Its effects are challenging to detect and counteract, necessitating vigilance in consuming information and seeking reputable sources (Bordignon and Pagano, 2022).

Fundamental research on disinformation focuses on its impact on democratic systems and the supply side of disinformation by looking at its producers. Disinformation campaigns could threaten democracy by destabilising democratic procedures and eroding public trust in institutions (Farkas and Schou, 2024). While experimental research suggests the limited direct impact on voter preferences, the broader societal implications, such as decreased trust and satisfaction with democratic institutions, are concerning (Koc-Michalska et al. 2020). Scholars describe this as disrupting the public sphere, harming the capacity for fact-based public debate (Bimber and Gil de Zúñiga 2020). Freedom House and V-Dem report highlighting how disinformation, polarisation, and automatisation reinforce each other, with top democratisers reducing disinformation and polarisation[8].

All these social and political dynamics have been accelerated and scaled up with the advent of AI, particularly generative AI. Chatbots, deepfakes, and sentiment analysis are increasingly used in online advocacy to enhance campaign efforts. Chatbots facilitate real-time interaction with supporters, providing instant responses and information dissemination. Sentiment analysis helps organisations gauge public opinion and adjust their strategies accordingly (Greco 2022).

AI has become a significant tool in election campaigns, and its use has become widespread since Barack Obama's presidential campaign in 2008. AI allows election teams to reach more of the population, as seen in India's 2024 election, where candidates' video messages were translated into multiple languages and dialects[9]. AI also enables more precise audience targeting, which is crucial in political campaigns, and can provide images and advertising for candidates at low cost or even for free.

However, while AI can improve campaign efficiency and personalisation, it also raises significant risks. AI-generated misinformation and ethical dilemmas are prevalent issues, with deepfakes being a notable example. These realistic but fake videos can manipulate public perception and spread false information. They can thus be used to promote unethical – if not illegal – competition between candidates. Recent research on deepfakes and cheapfakes highlights significant advancements and concerns in digital misinformation. Deepfakes use sophisticated AI to create highly realistic but fake videos, posing severe threats to information integrity and public trust. Conversely, cheapfakes, which are simpler to produce, involve basic editing techniques to manipulate content.

---

[8] https://v-dem.net/media/publications/PB39.pdf , accessed on 4 July 2024.
[9] https://www.techpolicy.press/the-era-of-aigenerated-election-campaigning-is-underway-in-india/ , accessed on 4 July 2024.

Studies indicate that both forms can significantly impact public opinion, political processes, and social stability by spreading false information (Paris and Donovan, 2019). For instance, in 2024 the US Republican National Committee (RNC) used an AI-generated video to criticise Joe Biden, exemplifying how AI can consolidate harmful competition and democratise disinformation processes in our societies. Also, during the 2024 French snap elections, the far-right parties National Rally (RN) and Reconquest have mobilised voters with scores of images and videos created mainly via MidJourney, a generative AI tool. In total, French political parties have collectively posted 23 AI-generated images in 81 posts across 81 posts on Facebook, Instagram and X. Almost all of the content promoted divisive images around immigration, Muslims, the European Union and French President Emmanuel Macron[10]. Despite techniques like watermarking, deepfakes remain a substantial regulatory challenge, complicating efforts to maintain information integrity (Malanowska et al. 2024).

AI can also be used to promote political falsehoods. For example, AI chatbots, such as OpenAI's ChatGPT, Microsoft's Bing Chat and Google's Bard, could be used by politicians to generate personalised campaign promises by deceptively targeting voters and donors (Kim and Lee 2023).

AI is crucial in processing and analysing large datasets for data-driven campaigns (TeBlunthuis 2022). AI algorithms analyse voter data to predict behaviour and tailor messages, significantly impacting campaign strategies. For example, during the 2019 Indian general election, AI-driven data analytics were used to craft highly targeted advertising campaigns, influencing voter decisions (Rathi 2019). AI's ability to predict voter behaviour and personalise outreach efforts enhances campaign efficiency. However, this power also raises concerns about the amplification of misinformation and the ethical use of AI for manipulation and propaganda. The potential for AI to spread false information and manipulate public opinion threatens democratic processes (Flynn et al. 2017). Addressing these ethical concerns is critical to ensuring the responsible use of AI in political campaigns.

Another concern involves AI and big data analysis for targeted advertising. For instance, during the 2021 federal election in Germany, the liberal party FDP deployed different advertisements based on user profiles. If a user's profile indicated an interest in sustainability, the FDP's ads promoted their climate measures. Conversely, if the profile suggested frequent travel, particularly by plane, their ads criticised red tape and opposed restrictions on unsustainable transport. These conflicting policy goals highlight inherent incompatibilities, leading to increased resentment as voters find that the outcomes do not match their expectations.

---

[10] https://aiforensics.org/work/french-elections-2024 (accessed on 4 July 2024).

All these uses and misuses of AI for political advocacy and campaigning are emerging in a very peculiar context. The attention economy, characterised by psychological and algorithmic manipulation on social media, plays a pivotal role in how AI affects online political advocacy and advertising. AI operates in a hybrid era where the distinction between war and peace is blurred. However, the real challenge lies not only in disinformation but also in covert, professionalised, and sustained information operations like FIMI. Moreover, the social media landscape is undergoing radical transformations, such as the "great decentralisation", a concept describing users migrating to different platforms based on their political beliefs. Simultaneously, platforms prioritise amplifying emotions through attention-driven algorithms, leading to hate-fuelled online conversations. Across societies worldwide, polarisation and mistrust have intensified.

## 6. Conclusions

Digital democracy, blending democratic processes with digital technologies, has evolved through electronic, virtual, and network democracy phases. These phases have expanded democratic engagement by leveraging technology for direct participation and global community building.

AI and social media offer both opportunities and challenges for digital democracy. They enhance political deliberation and participation by lowering information access and expression barriers. However, AI's potential to spread misinformation, invade privacy, and reinforce biases raises significant concerns. The individualisation of digital citizenship through AI can lead to stereotyping and hinder personal identity development.

Digital platforms have reshaped political participation, offering new civic engagement and advocacy avenues. AI enhances these processes through personalised communication, real-time monitoring, and data analysis but also poses risks of manipulation and disinformation.

AI improves efficiency and integrity in modern electoral processes through voter registration, e-voting, and result tabulation. However, it also raises privacy, security and trust issues. AI's predictive capabilities in electoral behaviour introduce new dynamics in political competition, raising ethical concerns about manipulation and democratic legitimacy.

Digital and AI technologies benefit the public sphere and political advocacy by democratising access and enhancing outreach. However, spreading misinformation and targeted propaganda risks negatively affect public discourse and democracy.

To ensure the mitigation of the primary AI and digital technology-related risks in digital democracy processes, enhancing citizens' digital media literacy is crucial. This includes educating individuals on critically assessing information, recognising misinformation, and understanding the underlying mechanisms of AI and digital platforms. On the supply side, robust AI and platform regulation must establish clear ethical guidelines and accountability measures. These regulations should prevent misuse, ensure transparency in AI-driven decision-making processes, and protect user privacy and data security. By addressing both the demand and supply sides, a more resilient and trustworthy digital democratic environment can be fostered.

## References

Arana-Catania, Miguel, Felix-Anselm Van Lier, Rob Procter, Nataliya Tkachenko, Yulan He, Arkaitz Zubiaga, and Maria Liakata. 2021. 'Citizen Participation and Machine Learning for a Better Democracy'. *Digital Government: Research and Practice* 2 (3): 27:1-27:22. https://doi.org/10.1145/3452118.

Baisnée, Olivier, Alizé Cavé, Cyriac Gousset, Jérémie Nollet, and Fanny Parent. 2022. 'The Digital Coverage of the Yellow Vest Movement as Protest Activity'. *French Politics* 20 (3): 529–49. https://doi.org/10.1057/s41253-022-00190-0.

Barberà, Oscar, Giulia Sandri, Patricia Correa, and Juan Rodríguez-Teruel. 2021. 'Political Parties Transition into the Digital Era'. In *Digital Parties: The Challenges of Online Organisation and Participation*, edited by Oscar Barberà, Giulia Sandri, Patricia Correa, and Juan Rodríguez-Teruel, 1–22. Cham: Springer International Publishing. https://doi.org/10.1007/978-3-030-78668-7_1.

Berg, Sebastian, and Jeanette Hofmann. 2021. 'Digital Democracy'. *Internet Policy Review: Journal on Internet Regulation* 10 (4): 1–23.

Borbáth, Endre, Swen Hutter, and Arndt Leininger. 2023. 'Cleavage Politics, Polarisation and Participation in Western Europe'. *West European Politics* 46 (4): 631–51. https://doi.org/10.1080/01402382.2022.2161786.

Boulianne, Shelley. 2022. 'Socially Mediated Political Consumerism'. *Information, Communication & Society* 25 (5): 609–17. https://doi.org/10.1080/1369118X.2021.2020872.

Buryakov, Daniil, Airo Hino, Mate Kovacs, and Uwe Serdült. 2022. 'Text Mining from Party Manifestos to Support the Design of Online Voting Advice Applications'. In *2022 9th International Conference on Behavioural and Social Computing (BESC)*, 1–7. https://doi.org/10.1109/BESC57393.2022.9995398.

Ceccarini, Luigi. 2021. *The Digital Citizen(Ship): Politics and Democracy in the Networked Society*. Edward Elgar Publishing.

Choi, Moonsun. 2016. 'A Concept Analysis of Digital Citizenship for Democratic Citizenship Education in the Internet Age'. *Theory & Research in*

Social Education 44 (4): 565–607. https://doi.org/10.1080/00933104.2016.1210549.

Cobbe, Jennifer. 2021. 'Algorithmic Censorship by Social Platforms: Power and Resistance'. *Philosophy & Technology* 34 (4): 739–66. https://doi.org/10.1007/s13347-020-00429-0.

Congge, Umar, María-Dolores Guillamón, Achmad Nurmandi, Salahudin, and Iradhad Taqwa Sihidi. 2023. 'Digital Democracy: A Systematic Literature Review'. *Frontiers in Political Science* 5 (February). https://doi.org/10.3389/fpos.2023.972802.

Curzon, James, Tracy Ann Kosa, Rajen Akalu, and Khalil El-Khatib. 2021. 'Privacy and Artificial Intelligence'. *IEEE Transactions on Artificial Intelligence* 2 (2): 96–108. https://doi.org/10.1109/TAI.2021.3088084.

Deth, Jan W van. 2014. 'A Conceptual Map of Political Participation'. *Acta Politica* 49 (3): 349–67. https://doi.org/10.1057/ap.2014.6.

Dzisah, Wilberforce S. 2018. 'Social Media and Elections in Ghana: Enhancing Democratic Participation'. *African Journalism Studies* 39 (1): 27–47. https://doi.org/10.1080/23743670.2018.1452774.

Earl, Jennifer, and Katrina Kimport. 2011. *Digitally Enabled Social Change: Activism in the Internet Age*. The MIT Press. https://doi.org/10.7551/mitpress/9780262015103.001.0001.

Ekman, Joakim, and Erik Amnå. 2012. 'Political Participation and Civic Engagement: Towards a New Typology'. *Human Affairs* 22 (3): 283–300. https://doi.org/10.2478/s13374-012-0024-1.

Feldstein, Steven. 2019. 'How Artificial Intelligence Is Reshaping Repression The Road to Digital Unfreedom'. *Journal of Democracy* 30 (1): 40–52.

———. 2021. *The Rise of Digital Repression: How Technology Is Reshaping Power, Politics, and Resistance*. Oxford University Press.

Gabriel, Saadia, Liang Lyu, James Siderius, Marzyeh Ghassemi, Jacob Andreas, and Asu Ozdaglar. 2024. 'Generative AI in the Era of "Alternative Facts"'. *An MIT Exploration of Generative AI*, March. https://doi.org/10.21428/e4baedd9.82175d26.

Garcí-a-Ull, Francisco-José, and Mónica Melero-Lázaro. 2023. 'Gender stereotypes in AI-generated images'. *Profesional de la información* 32 (5). https://doi.org/10.3145/epi.2023.sep.05.

Gemenis, Kostas. 2024. 'Artificial Intelligence and Voting Advice Applications'. *Frontiers in Political Science* 6 (January). https://doi.org/10.3389/fpos.2024.1286893.

Gillern, Sam von, Benjamin Gleason, and Amy Hutchison. 2022. 'Digital Citizenship, Media Literacy, and the ACTS Framework'. *The Reading Teacher* 76 (2): 145–58. https://doi.org/10.1002/trtr.2120.

Goldstein, Stéphane. 2020. *Informed Societies*. Facet Publishing.

Habermas, Jürgen. 1989. *The Structural Transformation of the Public Sphere: An Inquiry into a Category of Bourgeois Society*. Translated by Thomas Burger and Lawrence Kert. Studies in Contemporary German Social Thought. Cambridge, Mass.: MIT Press.

Habermas, Jürgen, Sara Lennox, and Frank Lennox. 1974. 'The Public Sphere: An Encyclopedia Article (1964)'. *New German Critique*, no. 3, 49–55. https://doi.org/10.2307/487737.

Hadfi, Rafik, Jawad Haqbeen, Sofia Sahab, and Takayuki Ito. 2021. 'Argumentative Conversational Agents for Online Discussions'. *Journal of Systems Science and Systems Engineering* 30 (4): 450–64. https://doi.org/10.1007/s11518-021-5497-1.

Jennings, Freddie J., Valeria P. Suzuki, and Alexis Hubbard. 2021. 'Social Media and Democracy: Fostering Political Deliberation and Participation'. *Western Journal of Communication* 85 (2): 147–67. https://doi.org/10.1080/10570314.2020.1728369.

Kawerau, Lukas, Nils B. Weidmann, and Alberto Dainotti. 2023. 'Attack or Block? Repertoires of Digital Censorship in Autocracies'. *Journal of Information Technology & Politics*, January. https://www.tandfonline.com/doi/abs/10.1080/19331681.2022.2037118.

Kelm, Ole, and Marco Dohle. 2018. 'Information, Communication and Political Consumerism: How (Online) Information and (Online) Communication Influence Boycotts and Buycotts'. *New Media & Society* 20 (4): 1523–42. https://doi.org/10.1177/1461444817699842.

Kim, Taekyung, Meng Hong Por, and Sung-Byung Yang. 2017. 'Winning the Crowd in Online Fundraising Platforms: The Roles of Founder and Project Features'. *Electronic Commerce Research and Applications* 25 (September):86–94. https://doi.org/10.1016/j.elerap.2017.09.002.

Koc-Michalska, Karolina, and Darren Lilleker. 2017. 'Digital Politics: Mobilization, Engagement, and Participation'. *Political Communication* 34 (1): 1–5. https://doi.org/10.1080/10584609.2016.1243178.

Kotek, Hadas, Rikker Dockum, and David Sun. 2023. 'Gender Bias and Stereotypes in Large Language Models'. In *Proceedings of The ACM Collective Intelligence Conference*, 12–24. CI '23. New York, NY, USA: Association for Computing Machinery. https://doi.org/10.1145/3582269.3615599.

Kymlicka, Will, and Wayne Norman. 2000. *Citizenship in Diverse Societies*. Oxford University Press. https://doi.org/10.1093/019829770X.001.0001.

Lee, Francis L.F., and Joseph Man Chan. 2016. 'Digital Media Activities and Mode of Participation in a Protest Campaign: A Study of the Umbrella Movement'. *Information, Communication & Society* 19 (1): 4–22. https://doi.org/10.1080/1369118X.2015.1093530.

Leibold, James. 2020. 'Surveillance in China's Xinjiang Region: Ethnic Sorting, Coercion, and Inducement'. *Journal of Contemporary China* 29 (121): 46–60. https://doi.org/10.1080/10670564.2019.1621529.

Leydet, Dominique. 2023. 'Citizenship'. In *The Stanford Encyclopedia of Philosophy*, edited by Edward N. Zalta and Uri Nodelman, Fall 2023. Metaphysics Research Lab, Stanford University.

https://plato.stanford.edu/archives/fall2023/entries/citizenship/.

Magnette, Paul. 2005. *Citizenship: The History of an Idea*. ECPR Press.

Mahesh, B., K. Pavan Kumar, Somula Ramasubbareddy, and E. Swetha. 2020. 'A Review on Data Deduplication Techniques in Cloud'. In *Embedded Systems and Artificial Intelligence*, edited by Vikrant Bhateja, Suresh Chandra Satapathy, and Hassan Satori, 825–33. Singapore: Springer. https://doi.org/10.1007/978-981-15-0947-6_78.

Manoharan, Aroon P., and Alex Ingrams. 2018. 'Conceptualizing E-Government from Local Government Perspectives'. *State and Local Government Review* 50 (1): 56–66. https://doi.org/10.1177/0160323X18763964.

Marcellis-Warin, Nathalie de, Frédéric Marty, Eva Thelisson, and Thierry Warin. 2022. 'Artificial Intelligence and Consumer Manipulations: From Consumer's Counter Algorithms to Firm's Self-Regulation Tools'. *AI and Ethics* 2 (2): 259–68. https://doi.org/10.1007/s43681-022-00149-5.

Martínez, Gonzalo, Lauren Watson, Pedro Reviriego, José Alberto Hernández, Marc Juarez, and Rik Sarkar. 2024. 'Towards Understanding the Interplay of Generative Artificial Intelligence and the Internet'. In *Epistemic Uncertainty in Artificial Intelligence*, edited by Fabio Cuzzolin and Maryam Sultana, 59–73. Cham: Springer Nature Switzerland. https://doi.org/10.1007/978-3-031-57963-9_5.

Monteith, Scott, Tasha Glenn, John R. Geddes, Peter C. Whybrow, Eric Achtyes, and Michael Bauer. 2024. 'Artificial Intelligence and Increasing Misinformation'. *The British Journal of Psychiatry* 224 (2): 33–35. https://doi.org/10.1192/bjp.2023.136.

Mosca, Lorenzo, and Davide Vittori. 2023. 'A Digital Principal? Substantive Representation in the Case of the Italian Five Star Movement'. *European Societies* 25 (4): 627–56. https://doi.org/10.1080/14616696.2022.2144638.

Mossberger, Karen, Caroline J. Tolbert, and Ramona S. Mcneal. 2007. *Digital Citizenship: The Internet, Society, and Participation*. MIT Press.

Nirban, Virendra Singh, Tanu Shukla, Partha Sarathi Purkayastha, Nachiket Kotalwar, and Labeeb Ahsan. 2023. 'The Role of AI in Combating Fake News and Misinformation'. In *Innovations in Bio-Inspired Computing and Applications*, edited by Ajith Abraham, Anu Bajaj, Niketa Gandhi, Ana Maria Madureira, and Cengiz Kahraman, 690–701. Cham: Springer Nature Switzerland. https://doi.org/10.1007/978-3-031-27499-2_64.

Norris, Pippa. 2002. *Democratic Phoenix: Reinventing Political Activism*. Cambridge: Cambridge University Press. https://doi.org/10.1017/CBO9780511610073.

Novelli, Claudio, Federico Casolari, Philipp Hacker, Giorgio Spedicato, and Luciano Floridi. 2024. 'Generative AI in EU Law: Liability, Privacy, Intellectual Property, and Cybersecurity'. SSRN Scholarly Paper. Rochester, NY. https://doi.org/10.2139/ssrn.4694565.

Novelli, Claudio, Giuliano Formisano, Prathm Juneja, Giulia Sandri, and Luciano Floridi. 2024. 'Artificial Intelligence for the Internal Democracy of Political Parties'. SSRN Scholarly Paper. Rochester, NY. https://doi.org/10.2139/ssrn.4778813.

Papakyriakopoulos, Orestis, Christelle Tessono, Arvind Narayanan, and Mihir Kshirsagar. 2022. 'How Algorithms Shape the Distribution of Political Advertising: Case Studies of Facebook, Google, and TikTok'. In *Proceedings of the 2022 AAAI/ACM Conference on AI, Ethics, and Society*, 532–46. AIES '22. New York, NY, USA: Association for Computing Machinery. https://doi.org/10.1145/3514094.3534166.

Ribble, Mike S., Gerald D. Bailey, and Tweed W. Ross. 2004. 'Digital Citizenship: Addressing Appropriate Technology Behavior'. *Learning & Leading with Technology* 32 (1): 6.

Sandri, Giulia, Fabio Garcia Lupato, Marco Meloni, Felix von Nostitz, and Oscar Barberà. 2024. 'Mapping the Digitalisation of European Political Parties'. *Information, Communication & Society* 0 (0): 1–22. https://doi.org/10.1080/1369118X.2024.2343369.

Saward, Michael. 2008. 'Democracy and Citizenship: Expanding Domains'. In *The Oxford Handbook of Political Theory*, edited by John S. Dryzek, Bonnie Honig, and Anne Phillips, 0. Oxford University Press. https://doi.org/10.1093/oxfordhb/9780199548439.003.0022.

Smith, Alan D., and John S. Clark. 2005. 'Revolutionising the Voting Process through Online Strategies'. *Online Information Review* 29 (5): 513–30. https://doi.org/10.1108/14684520510628909.

Sourati, Zhivar, Vishnu Priya Prasanna Venkatesh, Darshan Deshpande, Himanshu Rawlani, Filip Ilievski, Hông-Ân Sandlin, and Alain Mermoud. 2023. 'Robust and Explainable Identification of Logical Fallacies in Natural Language Arguments'. *Knowledge-Based Systems* 266 (April):110418. https://doi.org/10.1016/j.knosys.2023.110418.

Stoycheff, Elizabeth, G. Scott Burgess, and Maria Clara Martucci. 2020. 'Online Censorship and Digital Surveillance: The Relationship between Suppression Technologies and Democratization across Countries'. *Information, Communication & Society*, March. https://www.tandfonline.com/doi/abs/10.1080/1369118X.2018.1518472.

Tamilselvi, M., B. Manimaran, and Sanasam Chanu Inunganbi. 2023. 'Empirical Assessment of Artificial Intelligence Enabled Electronic Voting System Using Face Biometric Verification Strategy'. In *2023 Eighth International Conference on Science Technology Engineering and Mathematics (ICONSTEM)*, 1–7. https://doi.org/10.1109/ICONSTEM56934.2023.10142923.

Teorell, Jan, Mariano Torcal, and José Ramón Montero. 2007. 'Political Participation: Mapping the Terrain'. In *Citizenship and Involvement in European Democracies: A Comparative Perspective*, edited by Jan van Deth, José Ramon Montero, and Anders Westholm, 17:334–57. Routledge.

Theocharis, Yannis, Shelley Boulianne, Karolina Koc-Michalska, and Bruce

Bimber. 2023. 'Platform Affordances and Political Participation: How Social Media Reshape Political Engagement'. *West European Politics* 46 (4): 788–811. https://doi.org/10.1080/01402382.2022.2087410.

Tsahkna, Anna-Greta. 2013. 'E-Voting: Lessons from Estonia'. *European View* 12 (1): 59–66. https://doi.org/10.1007/s12290-013-0261-7.

Tsareva, Daniela. 2020. 'Electronic Voting in Bulgaria and Germany'. *Politics & Security* 4 (1): 87–109.

Verba, Sidney, and Norman H. Nie. 1987. *Participation in America: Political Democracy and Social Equality*. Chicago, IL: University of Chicago Press. https://press.uchicago.edu/ucp/books/book/chicago/P/bo3637096.html.

Verba, Sidney, Kay Lehman Schlozman, and Henry Brady. 1995. *Voice and Equality: Civic Voluntarism in American Politics*. Cambridge, Mass: Harvard University Press.

Wang, King-Hang, S. Mondal, Ki Chan, and Xuecai Xie. 2017. 'A Review of Contemporary E-Voting : Requirements , Technology , Systems and Usability'. 2017. https://www.semanticscholar.org/paper/A-Review-of-Contemporary-E-voting-%3A-Requirements-%2C-Wang-Mondal/e734d63888d81075efa0402599ae4e43772cf2e7.

Xu, Danni, Shaojing Fan, and Mohan Kankanhalli. 2023. 'Combating Misinformation in the Era of Generative AI Models'. In *Proceedings of the 31st ACM International Conference on Multimedia*, 9291–98. MM '23. New York, NY, USA: Association for Computing Machinery. https://doi.org/10.1145/3581783.3612704.

Zarnoufi, Randa, Mehdi Boutbi, and Mounia Abik. 2020. 'AI to Prevent Cyber-Violence: Harmful Behaviour Detection in Social Media'. *International Journal of High Performance Systems Architecture* 9 (4): 182–91. https://doi.org/10.1504/IJHPSA.2020.113679.

Zhang, Angie, Olympia Walker, Kaci Nguyen, Jiajun Dai, Anqing Chen, and Min Kyung Lee. 2023. 'Deliberating with AI: Improving Decision-Making for the Future through Participatory AI Design and Stakeholder Deliberation'. *Proceedings of the ACM on Human-Computer Interaction* 7 (CSCW1): 125:1-125:32. https://doi.org/10.1145/3579601.

Zhao, Yu-qing, Xiao-bo Yan, and Lin Wang. 2021. 'Ballot Character Recognition Based on Image Processing'. *Journal of Physics: Conference Series* 2024 (1): 012010. https://doi.org/10.1088/1742-6596/2024/1/012010.